# Simple realization of a hybrid controlled–controlled-Z gate with photonic control qubits encoded via eigenstates of the photon-number parity operator

Qi-Ping Su,[1] Liang Bin,[1] Yu Zhang,[2] and Chui-Ping Yang[1,a]

**AFFILIATIONS**

[1]School of Physics, Hangzhou Normal University, Hangzhou 311121, China
[2]School of Physics, Nanjing University, Nanjing 210093, China

[a]Author to whom correspondence should be addressed: yangcp@hznu.edu.cn

**ABSTRACT**

We propose a simple method to realize a hybrid controlled–controlled-Z (CCZ) gate with two photonic qubits simultaneously controlling a superconducting (SC) target qubit, by employing two microwave cavities coupled to a SC ququart (a four-level quantum system). In this proposal, each control qubit is a photonic qubit, which is encoded by two *arbitrary* orthogonal eigenstates (with eigenvalues ±1, respectively) of the photon-number parity operator. Since the two *arbitrary* encoding states can take various quantum states, this proposal can be applied to realize the hybrid CCZ gate, for which the two control photonic qubits can have various encodings. The gate realization is quite simple because only a basic operation is needed. During the gate operation, the higher energy intermediate levels of the ququart are not occupied, and, thus, decoherence from these levels is greatly suppressed. We further discuss how to apply this gate to generate a hybrid Greenberger–Horne–Zeilinger (GHZ) entangled state of a SC qubit and two photonic qubits, which takes a general form. As an example, our numerical simulation demonstrates that high-fidelity generation of a cat–cat–spin hybrid GHZ state is feasible within current circuit QED technology. This proposal is quite general, which can be applied to realize the hybrid CCZ gate as well as to prepare various hybrid GHZ states of a matter qubit and two photonic qubits in other physical systems, such as two microwave or optical cavities coupled to a four-level natural or artificial atom.

Multiquit gates play important roles in quantum computing. Especially, a controlled–controlled-Z gate (CCZ gate) with two qubits simultaneously controlling a target qubit is of significance, which has practical applications in quantum computing, such as quantum circuit construction, error correction, and quantum algorithms.[1–5] Experimentally, a non-hybrid CCZ gate with three matter qubits was demonstrated in various physical systems.[6–9] On the other hand, hybrid gates acting on "different types of qubits" have attracted increasing attention due to their important applications in hybrid quantum computing. Here, different types of qubits means: qubits are different in their nature (e.g., photonic qubits and matter qubits) or different in their encoding (e.g., encoding through discrete variables and encoding via continuous variables), etc. The focus of this work is on a hybrid CCZ gate with two photonic qubits simultaneously controlling a target superconducting (SC) qubit. Since a photonic qubit is different from a SC qubit in their nature, the CCZ gate considered in this work is a hybrid CCZ gate. Obviously, a hybrid CCZ gate differs from a regular non-hybrid CCZ gate with three identical qubits.

Over the past years, proposals have been put forward for implementing a hybrid two-qubit controlled phase or NOT gate with (i) a matter qubit (e.g., SC qubit, nitrogen-vacancy (NV)-center qubit, atomic qubit, quantum-dot qubit, electron-spin qubit, etc.) and a photonic qubit[10–17] or (ii) two photonic qubits with different encodings.[18–20] In addition, proposals have been presented for implementing a hybrid Toffoli gate or a hybrid CCZ gate with (i) two NV-center qubits and a photonic qubit[11] or (ii) two quantum-dot qubits and a photonic qubit.[17] Moreover, a hybrid Fredkin gate, with two quantum-dot qubits and a photonic qubit, has been presented.[17] We note that, in the previous proposals,[10–20] photonic qubits are encoded via cat states, coherent states, polarization states, spatial mode degrees of freedom, or the vacuum and the single-photon states. After a deep search of literature, we find that how to realize a hybrid CCZ gate, with two photonic qubits (encoded by two arbitrary orthogonal eigenstates of the photon-number parity operator) simultaneously controlling a target matter qubit (e.g., SC qubit or other matter qubit), has not been reported to date.

In this work, we will present a simple method to *directly* realize a hybrid CCZ gate with two photonic qubits simultaneously controlling a SC target qubit, by using two microwave cavities coupled to a SC flux ququart (a four-level artificial atom) [Fig. 1(a)]. Each cavity can be a one-dimensional (1D) or three-dimensional (3D) cavity. For each cavity being a 3D cavity, the ququart is inductively coupled to the cavity, which can be implemented by inserting the part of the superconducting loop of the SC ququard into the cavity.[21] While, for each cavity being a 1D cavity, the ququard is capacitively coupled to the cavity, which can be realized by using a capacitor to connect the ququard and the cavity. Note that for either of these two couplings, the key Hamiltonian of this work, as described by Eq. (3), can be obtained.

This proposal is based on circuit QED, which has been considered as one of the best platforms for quantum computing.[22–27] In this proposal, the two logic states of each photonic qubit are encoded by two *arbitrary* orthogonal eigenstates $|\varphi_e\rangle$ and $|\varphi_o\rangle$ of the photon-number parity operator $\hat{\pi} = e^{i\pi \hat{a}^+ \hat{a}}$ of a cavity. Here, $\hat{a}$ ($\hat{a}^+$) is the photon-annihilation (creation) operator. In addition, for the SC target qubit, the two logic states are encoded by the two lowest levels $|g'\rangle$ and $|g\rangle$ of a SC ququart [Fig. 1(b)]. As shown below, two hybrid two-body interactions are simultaneously turned onto produce the three-body interaction that ends up producing the desired hybrid CCZ. We should mention that this idea has also been applied in superconducting systems to realize non-hybrid three-qubit gates with SC qubits.[28–30]

The two arbitrary encoding states $|\varphi_e\rangle$ and $|\varphi_o\rangle$ can be expressed as

$$|\varphi_e\rangle = \sum_{m=0}^{\infty} C_{2m}|2m\rangle, \quad |\varphi_o\rangle = \sum_{n=0}^{\infty} C_{2n+1}|2n+1\rangle, \quad (1)$$

where $m$ and $n$ are non-negative integers, and the coefficients $C_{2m}$ and $C_{2n+1}$ satisfy the normalization conditions. Obviously, the two states $|\varphi_e\rangle$ and $|\varphi_o\rangle$ are orthogonal to each other. One can easily check $\hat{\pi}|\varphi_e\rangle = |\varphi_e\rangle$ and $\hat{\pi}|\varphi_o\rangle = -|\varphi_o\rangle$, namely, the two encoding states $|\varphi_e\rangle$ and $|\varphi_o\rangle$ are the eigenstates of the photon-number parity operator $\hat{\pi}$ with eigenvalues 1 and −1, respectively.

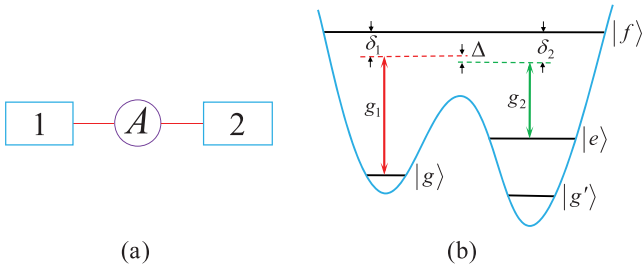

**FIG. 1.** (a) Schematic circuit of two microwave cavities coupled to a SC flux ququart. Each square represents a 1D or 3D microwave cavity. The circle $A$ represents the flux ququart, which is inductively or capacitively coupled to each cavity. (b) Cavity 1 is dispersively coupled to the $|g\rangle \leftrightarrow |f\rangle$ transition of the ququart with coupling strength $g_1$ and detuning $\delta_1$, while cavity 2 is dispersively coupled to the $|e\rangle \leftrightarrow |f\rangle$ transition of the ququart with coupling strength $g_2$ and detuning $\delta_2$. Here, $\Delta = \delta_2 - \delta_1$.

For the hybrid CCZ gate here, there are a total number of eight computational basis states, denoted by $|\varphi_e\rangle|\varphi_e\rangle|g'\rangle$, $|\varphi_e\rangle|\varphi_e\rangle|g\rangle$, $|\varphi_e\rangle|\varphi_o\rangle|g'\rangle$, $|\varphi_e\rangle|\varphi_o\rangle|g\rangle$, $|\varphi_o\rangle|\varphi_e\rangle|g'\rangle$, $|\varphi_o\rangle|\varphi_e\rangle|g\rangle$, $|\varphi_o\rangle|\varphi_o\rangle|g'\rangle$, and $|\varphi_o\rangle|\varphi_o\rangle|g\rangle$. The hybrid CCZ gate is described by the following state transformations:

$$\begin{aligned} |\varphi_o\rangle|\varphi_o\rangle|g\rangle &\rightarrow -|\varphi_o\rangle|\varphi_o\rangle|g\rangle, \\ |l_1 l_2 l_3\rangle &\rightarrow |l_1 l_2 l_3\rangle, \end{aligned} \quad (2)$$

where $|l_1 l_2 l_3\rangle = |\varphi_e\rangle|\varphi_e\rangle|g'\rangle$, $|\varphi_e\rangle|\varphi_e\rangle|g\rangle$, $|\varphi_e\rangle|\varphi_o\rangle|g'\rangle$, $|\varphi_e\rangle|\varphi_o\rangle|g\rangle$, $|\varphi_o\rangle|\varphi_e\rangle|g'\rangle$, $|\varphi_o\rangle|\varphi_e\rangle|g\rangle$, or $|\varphi_o\rangle|\varphi_o\rangle|g'\rangle$, which shows that when the two control photonic qubits are in the state $|\varphi_o\rangle$, a phase flip happens to the state $|g\rangle$ of the SC target qubit; when even one of the photonic qubits is not in the state $|\varphi_o\rangle$, nothing happens to the states $|g'\rangle$ and $|g\rangle$ of the SC target qubit.

Both photonic qubits and SC qubits have been considered as promising qubits and have been widely used in quantum computing. The hybrid CCZ gate considered here is important in hybrid quantum computing, such as hybrid error correction and hybrid quantum algorithms involving photonic qubits and SC qubits. Such hybrid gate is significant for implementing large-scale hybrid quantum computing performed in a compounded information processor, which consists of SC-qubit-based quantum processors and photonic-qubit-based quantum processors. Moreover, this hybrid gate is very important in the context of quantum state transfer between a SC-qubit-based quantum processor and a photonic-qubit-based quantum memory. The architecture consisting of a SC processor and a quantum memory has been shown to provide a significant interest.[31–33]

We should mention that the hybrid CCZ gate can, in principle, be constructed using basic two-qubit gates and single-qubit gates only. However, when using the gate-decomposing protocols, five two-qubit gates[34] or three two-qubit gates plus two single-qubit gates[35] are required to construct a CCZ gate. Therefore, building the hybrid CCZ gate may become complex since each elementary gate requires turning on and off a given Hamiltonian for a certain period of time, and each additional basic gate adds experimental complications and the possibility of more errors. However, by using the present proposal, implementing the hybrid CCZ gate is greatly simplified because it requires only a basic operation.

Our proposal also provides a simple way to realize a hybrid controlled–controlled-NOT gate (Toffoli gate) with the proposed two photonic qubits simultaneously controlling a SC target qubit, since a Toffoli gate can be constructed by a CCZ gate plus two single-qubit Hadamard gates, which are, respectively, performed on the target qubit before and after the CCZ gate.[1]

The four levels of the SC ququart are labeled as $|g'\rangle$, $|g\rangle$, $|e\rangle$, and $|f\rangle$ [Fig. 1(b)]. The $|g'\rangle \leftrightarrow |g\rangle$ transition can be made weak by increasing the barrier between two potential wells. Such a four-level SC ququart can be experimentally engineered by using a three-junction flux system with the Hamiltonian described in Ref. 36. The energies and detunings of the four levels can be accurately controlled in experiment.[37,38]

The ququart is initially decoupled from the two cavities. Now adjust the level spacings of the ququart or the frequency of each cavity, such that cavity 1 is dispersively coupled to the $|g\rangle \leftrightarrow |e\rangle$ transition with coupling constant $g_1$ and detuning $\delta_1$, and cavity 2 is dispersively coupled to the $|e\rangle \leftrightarrow |f\rangle$ transition with coupling constant $g_2$ and detuning $\delta_2$ [Fig. 1(b)]. Note that for a SC quantum device, the level spacings can be rapidly (1 − 3 ns) adjusted via changing by changing external control parameters.[32] In addition, the frequency of a microwave cavity or resonator can be quickly tuned within a few nanoseconds.[33]

In the interaction picture and after making the rotating-wave approximation (RWA), the Hamiltonian of the whole system can be written as (hereafter assuming $\hbar = 1$)

$$H_I = g_1 e^{-i\delta_1 t} \hat{a}_1^+ \sigma_{fg}^- + g_2 e^{-i\delta_2 t} \hat{a}_2^+ \sigma_{fe}^- + \text{H.c.}, \quad (3)$$

where $\hat{a}_1$ ($\hat{a}_2$) is the photon-annihilation operator of cavity 1 (2), $\sigma_{fg}^- = |g\rangle\langle f|$, $\sigma_{fe}^- = |e\rangle\langle f|$, $\delta_1 = \omega_{fg} - \omega_{c_1} > 0$, and $\delta_2 = \omega_{fe} - \omega_{c_2} > 0$ [Fig. 1(b)]. Here, $\omega_{fg}$ ($\omega_{fe}$) is the $|f\rangle \leftrightarrow |g\rangle$ ($|f\rangle \leftrightarrow |e\rangle$) transition frequency of the ququart, while $\omega_{c_1}$ ($\omega_{c_2}$) is the frequency of cavity 1 (2).

For the large-detuning condition $\delta_1 \gg g_1$ and $\delta_2 \gg g_2$, the Hamiltonian (3) becomes[39–41] (see the supplementary material)

$$H_e = -\lambda_1 \hat{n}_1 |g\rangle\langle g| + \lambda_1 (1 + \hat{n}_1)|f\rangle\langle f| - \lambda_2 \hat{n}_2 |e\rangle\langle e|$$
$$+ \lambda_2 (1 + \hat{n}_2)|f\rangle\langle f| - \lambda(e^{i\Delta t} \hat{a}_1^+ \hat{a}_2 \sigma_{eg}^- + \text{H.c.}), \quad (4)$$

where $\sigma_{eg}^- = |g\rangle\langle e|$, $\lambda_1 = g_1^2/\delta_1$, $\lambda_2 = g_2^2/\delta_2$, $\lambda = (g_1 g_2/2)(1/\delta_1 + 1/\delta_2)$, and $\Delta = \delta_2 - \delta_1 = \omega_{c_1} - \omega_{c_2} - \omega_{eg} > 0$. For $\Delta \gg \{\lambda_1, \lambda_l, \lambda\}$, the effective Hamiltonian $H_e$ turns into[39–41] (see the supplementary material)

$$H_e = -\lambda_1 \hat{n}_1 |g\rangle\langle g| + \lambda_1 (1 + \hat{n}_1)|f\rangle\langle f| - \lambda_2 \hat{n}_2 |e\rangle\langle e|$$
$$+ \lambda_2 (1 + \hat{n}_2)|f\rangle\langle f| + \chi \hat{n}_1 (1 + \hat{n}_2)|g\rangle\langle g| - \chi(1 + \hat{n}_1)\hat{n}_2 |e\rangle\langle e|, \quad (5)$$

where $\chi = \lambda^2/\Delta$. When the levels $|e\rangle$ and $|f\rangle$ are initially not occupied, these levels will remain unpopulated because the Hamiltonian (5) does not induce the $|g\rangle \to |e\rangle$ transition or the $|g\rangle \to |f\rangle$ transition. In this case, the effective Hamiltonian (5) reduces to

$$H_e = \eta \hat{n}_1 |g\rangle\langle g| + \chi \hat{n}_1 \hat{n}_2 |g\rangle\langle g|, \quad (6)$$

where $\eta = -\lambda_1 + \chi$.

Under the Hamiltonian (6), the unitary operator $U = e^{-iH_e t}$, describing the state time evolution of the system, can be expressed as

$$U = \exp(-i\eta \hat{n}_1 |g\rangle\langle g| t) \exp(-i\chi \hat{n}_1 \hat{n}_2 |g\rangle\langle g| t). \quad (7)$$

According to the encoding in Eq. (1), the unitary operation $U$ results in the following state transformation:

$$U|\varphi_e\rangle|\varphi_e\rangle|g\rangle = \sum_{m,m'} A_{mm'} C_{2m} C_{2m'} |2m\rangle|2m'\rangle|g\rangle,$$
$$U|\varphi_e\rangle|\varphi_o\rangle|g\rangle = \sum_{m,n'} A_{mn'} C_{2m} C_{2n'+1} |2m\rangle|2n'+1\rangle|g\rangle,$$
$$U|\varphi_o\rangle|\varphi_e\rangle|g\rangle = \sum_{n,m'} A_{nm'} C_{2n+1} C_{2m'} |2n+1\rangle|2m'\rangle|g\rangle,$$
$$U|\varphi_o\rangle|\varphi_o\rangle|g\rangle = \sum_{n,n'} A_{nn'} C_{2n+1} C_{2n'+1} |2n+1\rangle|2n'+1\rangle|g\rangle, \quad (8)$$

where $m$ and $n$ are associated with the first photonic qubit, while $m'$ and $n'$ are associated with the second photonic qubit, $A_{mm'} = \exp(-i2m\eta t)\exp(-i2m \times 2m' \chi t)$, $A_{mn'} = \exp(-i2m\eta t)\exp[-i2m \times (2n'+1)\chi t]$, $A_{nm'} = \exp[-i(2n+1)\eta t]\exp[-i(2n+1)2m' \chi t]$, and $A_{nn'} = \exp[-i(2n+1)\eta t]\exp[-i(2n+1)(2n'+1)\chi t]$. For $\chi t = \pi$ and $\eta t = 2s\pi$ ($s$ is an integer), one has $A_{mm'} = A_{mn'} = A_{nm'} = 1$ and $A_{nn'} = -1$. Thus, the state transformation (8) becomes

$$U|\varphi_e\rangle|\varphi_e\rangle|g\rangle = |\varphi_e\rangle|\varphi_e\rangle|g\rangle,$$
$$U|\varphi_e\rangle|\varphi_o\rangle|g\rangle = |\varphi_e\rangle|\varphi_o\rangle|g\rangle,$$
$$U|\varphi_o\rangle|\varphi_e\rangle|g\rangle = |\varphi_o\rangle|\varphi_e\rangle|g\rangle,$$
$$U|\varphi_o\rangle|\varphi_o\rangle|g\rangle = -|\varphi_o\rangle|\varphi_o\rangle|g\rangle, \quad (9)$$

which indicates that when the two control photonic qubits are in the state $|\varphi_o\rangle$, a phase flip (from sign + to −) happens to the state $|g\rangle$ of the SC target qubit.

On the other hand, since the level $|g'\rangle$ is not involved in the unitary operator $U$, the four basic states $|\varphi_e\rangle|\varphi_e\rangle|g'\rangle$, $|\varphi_e\rangle|\varphi_o\rangle|g'\rangle$, $|\varphi_o\rangle|\varphi_e\rangle|g'\rangle$, and $|\varphi_o\rangle|\varphi_o\rangle|g'\rangle$ remain unchanged. Hence, it can be concluded from Eq. (9) that, after the above operation, the hybrid CCZ gate (2) is realized. After the gate operation, one needs to adjust the level spacings of the ququart or the frequency of each cavity, such that the ququart is decoupled from the two cavities.

From the description presented above, one can clearly see that

(i) The hybrid CCZ gate is implemented through a basic operation described by the unitary operator $U$.
(ii) The higher energy intermediate levels $|e\rangle$ and $|f\rangle$ of the ququart are not occupied during the operation.
(iii) We have set $\chi t = \pi$, $\eta t = 2s\pi$, and $\eta = -\lambda_1 + \chi$, resulting in $-\lambda_1 t + \pi = 2s\pi$, which can be met by choosing $\lambda_1 t = (2k+1)\pi$. Combining $\chi t = \pi$ and $\lambda_1 t = (2k+1)\pi$ yields

$$g_2 = \frac{2\delta_2}{\delta_1 + \delta_2}\sqrt{\Delta \delta_1/(2k+1)}. \quad (10)$$

This condition can be readily achieved by adjusting $\delta_2$, given $\delta_1$. Because of $\delta_2 = \omega_{fe} - \omega_{c_2}$, the detuning $\delta_2$ can be adjusted by varying the frequency $\omega_{c_2}$ of cavity 2.

Note that the two arbitrary encoding states $|\varphi_e\rangle$ and $|\varphi_o\rangle$ can be various specific quantum states. For instances, they can be (i) $|\varphi_e\rangle = |0\rangle$ (vacuum state), $|\varphi_o\rangle = |1\rangle$ (single-photon state); (ii) $|\varphi_e\rangle = |cat\rangle$, $|\varphi_o\rangle = |\overline{cat}\rangle$ (cat state), here $|cat\rangle = \mathcal{N}(|\alpha\rangle + |-\alpha\rangle)$ and $|\overline{cat}\rangle = \mathcal{N}(|\alpha\rangle - |-\alpha\rangle)$, with a normalization factor $\mathcal{N}$; (iii) $|\varphi_e\rangle = |2m\rangle$ (Fock state with even-number photons), $|\varphi_o\rangle = |2n+1\rangle$ (Fock state with odd-number photons), and so on. Therefore, this proposal can be used to realize the hybrid CCZ gate (2), for which the two control photonic qubits can have various encodings.

Hybrid entangled states can serve as quantum channels and intermediate resources for various quantum tasks, covering the transmission, operation, and storage of quantum information between different formats and encodings.[42,43] As an application, we will show how to apply the hybrid CCZ gate to create a hybrid Greenberger–Horne–Zeilinger (GHZ) entangled state of a SC qubit and two photonic qubits, which takes a general form.

Let us go back to the physical system illustrated in Fig. 1(a). Assume that the coupler SC ququart is initially in the state $(|g'\rangle + |g\rangle)/\sqrt{2}$, and the two cavities are initially prepared in the entangled state $(|\varphi_e\rangle|\varphi_e\rangle + |\varphi_o\rangle|\varphi_o\rangle)/\sqrt{2}$. Note that the procedure or the method, used for the initial state preparation of the two cavities, strongly depends on what the two encoding states $|\varphi_e\rangle$ and $|\varphi_o\rangle$ are. To generate the entangled state of the two cavities, the coupling between the two cavities is required. In principle, such a coupling does not affect the gate performance because it can be fully tuned on or off before the gate operation. The cavity–cavity coupling can be indirectly tuned on by the direct coupling of the SC ququart with each cavity.[44,45] It can also be tuned off by the decoupling of the SC ququart from each cavity. As mentioned previously, the coupling or decoupling of the SC ququart with each cavity can be achieved by adjusting the level spacings of the SC ququart[32] or the frequency of each cavity.[33]

The initial state of the whole system is given by

$$|\psi(0)\rangle = \frac{1}{\sqrt{2}}(|\varphi_e\rangle|\varphi_e\rangle + |\varphi_o\rangle|\varphi_o\rangle)|+\rangle, \quad (11)$$

where $|+\rangle = (|g'\rangle + |g\rangle)/\sqrt{2}$. By applying the hybrid CCZ gate (2), the state (11) becomes

$$|GHZ\rangle = \frac{1}{\sqrt{2}}(|\varphi_e\rangle|\varphi_e\rangle|+\rangle + |\varphi_o\rangle|\varphi_o\rangle|-\rangle), \quad (12)$$

where $|-\rangle = (|g'\rangle - |g\rangle)/\sqrt{2}$. Here, the two logic states of the SC qubit are encoded via the two rotated spin states $|+\rangle$ and $|-\rangle$ of the SC ququart. The state (12) is a hybrid GHZ entangled state of a SC qubit and two photonic qubits, which takes a general form. From Eq. (12), one can see that, depending on the specific encodings of $|\varphi_e\rangle$ and $|\varphi_o\rangle$, various hybrid GHZ states can be prepared.

As an example, we now investigate the experimental feasibility for creating a cat-cat–spin hybrid GHZ state,

$$|GHZ\rangle = \frac{1}{\sqrt{2}}(|cat\rangle|cat\rangle|+\rangle + |\overline{cat}\rangle|\overline{cat}\rangle|-\rangle). \quad (13)$$

Based on Eq. (12), this hybrid GHZ state is obtained for the photonic qubit encoding $|\varphi_e\rangle = |cat\rangle$ and $|\varphi_o\rangle = |\overline{cat}\rangle$.

The physical system, used for the GHZ state generation, consists of two 1D microwave cavities coupled to a SC flux ququart (Fig. 2). According to Eq. (11), the required initial state of the system is

$$|\psi(0)\rangle = \frac{1}{\sqrt{2}}(|cat\rangle|cat\rangle + |\overline{cat}\rangle|\overline{cat}\rangle)|+\rangle. \quad (14)$$

This initial state is available in experiments because the entangled cat state $(|cat\rangle|cat\rangle + |\overline{cat}\rangle|\overline{cat}\rangle)/\sqrt{2}$ of two microwave cavities was generated in the circuit QED experiments,[46] and the state $|+\rangle$ can be easily prepared by applying a classical pulse resonant with the $|g'\rangle \leftrightarrow |g\rangle$ transition of the SC ququart in the ground state $|g'\rangle$.

In reality, there exist the unwanted couplings of each cavity with the SC ququart and the unwanted inter-cavity crosstalk. When they are considered, the Hamiltonian (3) is modified as

$$H'_I = g_1 e^{-i\delta_1 t} \hat{a}_1^+ \sigma_{fg}^- + g_2 e^{-i\delta_2 t} \hat{a}_2^+ \sigma_{fe}^- + g'_1 e^{-i\delta'_1 t} \hat{a}_1^+ \sigma_{fe}^-$$
$$+ g''_1 e^{-i\delta''_1 t} \hat{a}_1^+ \sigma_{eg}^- + g'''_1 e^{-i\delta'''_1 t} \hat{a}_1^+ \sigma_{eg'}^- + g'_2 e^{-i\delta'_2 t} \hat{a}_2^+ \sigma_{fg}^-$$
$$+ g''_2 e^{-i\delta''_2 t} \hat{a}_2^+ \sigma_{eg}^- + g'''_2 e^{-i\delta'''_2 t} \hat{a}_2^+ \sigma_{eg'}^- + g_{12} e^{i\Delta_{12} t} \hat{a}_1^+ \hat{a}_2 + \text{H.c.}, \quad (15)$$

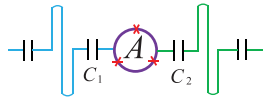

FIG. 2. Setup for two microwave cavities and a SC flux ququart (the circle A in the middle). Each cavity is a 1D transmission line resonator. The flux ququart, consisting of three Josephson junctions and a superconducting loop, is connected to each cavity via a capacitor.

where $\sigma_{eg}^- = |g\rangle\langle e|, \sigma_{eg'}^- = |g'\rangle\langle e|; g'_i, g''_i$, and $g'''_i$ are the coupling constants of cavity $i$ with the corresponding inter-level transitions of the SC ququart ($i = 1, 2$) (Fig. 3); the detunings are defined as $\delta'_i = \omega_{fe} - \omega_{c_i}, \delta''_i = \omega_{fg} - \omega_{c_i}, \delta''_i = \omega_{eg} - \omega_{c_i}$, and $\delta'''_i = \omega_{eg'} - \omega_{c_i}$ ($i = 1, 2$) (Fig. 3); $g_{12}$ is the two-cavity crosstalk strength, while $\Delta_{12} = \omega_{c_1} - \omega_{c_2}$ is the frequency detuning of the two cavities. Here, $\omega_{eg}$ ($\omega_{eg'}$) is the $|g\rangle \leftrightarrow |e\rangle$ ($|g'\rangle \leftrightarrow |e\rangle$) transition frequency of the SC ququart. Note that the coupling of each cavity with the $|g'\rangle \leftrightarrow |g\rangle$ and $|g'\rangle \leftrightarrow |f\rangle$ transitions can be neglected because of the weak $|g'\rangle \leftrightarrow |g\rangle$ transition or by adjusting the level spacings of the SC ququart, such that each cavity is highly detuned (decoupled) from these transitions.

The dynamics of the lossy system is governed by the master equation

$$\frac{d\rho}{dt} = -i[H'_I, \rho] + \kappa_1 \mathcal{L}[\hat{a}_1] + \kappa_2 \mathcal{L}[\hat{a}_2] + \sum_{fe} \gamma_{fe} \mathcal{L}[\sigma_{fe}^-]$$
$$+ \sum_{fg} \gamma_{fg} \mathcal{L}[\sigma_{fg}^-] + \sum_{fg'} \gamma_{fg'} \mathcal{L}[\sigma_{fg'}^-] + \sum_{eg} \gamma_{eg} \mathcal{L}[\sigma_{eg}^-]$$
$$+ \sum_{eg'} \gamma_{eg'} \mathcal{L}[\sigma_{eg'}^-] + \sum_{gg'} \gamma_{gg'} \mathcal{L}[\sigma_{gg'}^-]$$
$$+ \sum_{s=f,e,g} \gamma_{s,\varphi} (\sigma_{ss}\rho\sigma_{ss} - \sigma_{ss}\rho/2 - \rho\sigma_{ss}/2), \quad (16)$$

where $\sigma_{ss} = |s\rangle\langle s|$ ($s = f, e, g$), $\mathcal{L}[\Lambda] = \Lambda\rho\Lambda^+ - \Lambda^+\Lambda\rho/2 - \rho\Lambda^+\Lambda/2$ ($\Lambda = \hat{a}_1, \hat{a}_2, \sigma_{fe}^-, \sigma_{fg}^-, \sigma_{fg'}^-, \sigma_{eg}^-, \sigma_{eg'}^-$); $\gamma_{fe}, \gamma_{fg}$, and $\gamma_{fg'}$ are the relaxation rates of the level $|f\rangle$ for the decay paths $|f\rangle \to |e\rangle, |f\rangle \to |g\rangle$, and $|f\rangle \to |g'\rangle$, respectively; $\gamma_{eg}$ ($\gamma_{eg'}$) is the energy relaxation rate of the level $|e\rangle$ for the decay path $|e\rangle \to |g\rangle$ ($|e\rangle \to |g'\rangle$), and $\gamma_{gg'}$ is the energy relaxation rate of the level $|g\rangle$; $\gamma_{f,\varphi}, \gamma_{e,\varphi}$, and $\gamma_{g,\varphi}$ are the dephasing rates of the levels $|f\rangle, |e\rangle$, and $|g\rangle$, respectively; $\kappa_1$ ($\kappa_2$) is the decay rate of cavity 1 (2).

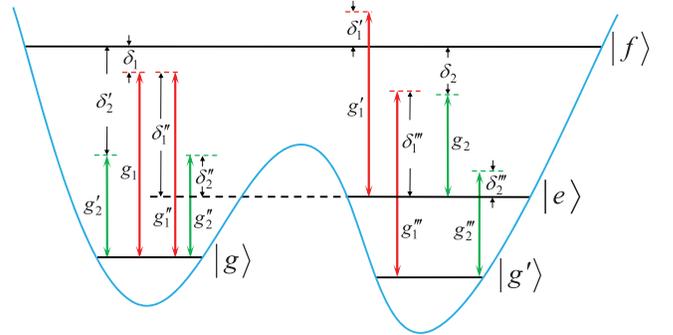

FIG. 3. Illustration of the dispersive coupling between cavity 1 and the $|g\rangle \leftrightarrow |f\rangle$ transition (with coupling constant $g_1$ and detuning $\delta_1$), the unwanted coupling between cavity 1 and the $|e\rangle \leftrightarrow |f\rangle$ transition (with coupling constant $g'_1$ and detuning $\delta'_1$), the unwanted coupling between cavity 1 and the $|g\rangle \leftrightarrow |e\rangle$ transition (with coupling constant $g''_1$ and detuning $\delta''_1$), and the unwanted coupling between cavity 1 and the $|g'\rangle \leftrightarrow |e\rangle$ transition (with coupling constant $g'''_1$ and detuning $\delta'''_1$). Illustration of the dispersive coupling between cavity 2 and the $|e\rangle \leftrightarrow |f\rangle$ transition (with coupling constant $g_2$ and detuning $\delta_2$), the unwanted coupling between cavity 2 and the $|g\rangle \leftrightarrow |f\rangle$ transition (with coupling constant $g'_2$ and detuning $\delta'_2$), the unwanted coupling between cavity 2 and the $|g\rangle \leftrightarrow |e\rangle$ transition (with coupling constant $g''_2$ and detuning $\delta''_2$), and the unwanted coupling between cavity 2 and the $|g'\rangle \leftrightarrow |e\rangle$ transition (with coupling constant $g'''_2$ and detuning $\delta'''_2$). Red lines correspond to cavity 1, while green lines correspond to cavity 2.

TABLE I. Parameters used in the numerical simulation.

| | | |
|---|---|---|
| $\omega_{gg'}/(2\pi) = 1.0$ GHz | $\omega_{eg}/(2\pi) = 7.0$ GHz | $\omega_{eg'}/(2\pi) = 8.0$ GHz |
| $\omega_{fe}/(2\pi) = 12.0$ GHz | $\omega_{fg}/(2\pi) = 19.0$ GHz | $\omega_{c_1}/(2\pi) = 18.3$ GHz |
| $\omega_{c_2}/(2\pi) = 11.2$ GHz | $\delta_1/(2\pi) = 0.7$ GHz | $\delta'_1/(2\pi) = -6.3$ GHz |
| $\delta''_1/(2\pi) = -11.3$ GHz | $\delta'''_1/(2\pi) = -10.3$ GHz | $\delta_2/(2\pi) = 0.8$ GHz |
| $\delta'_2/(2\pi) = 7.8$ GHz | $\delta''_2/(2\pi) = -4.2$ GHz | $\delta'''_2/(2\pi) = -3.2$ GHz |
| $\Delta_{12}/(2\pi) = 7.1$ GHz | $g_1/(2\pi) = 95.7$ MHz | $g'_1/(2\pi) = 95.7$ MHz |
| $g''_1/(2\pi) = 9.57$ MHz | $g'''_1/(2\pi) = 67.7$ MHz | $g_2/(2\pi) = 85.1$ MHz |
| $g'_2/(2\pi) = 85.1$ MHz | $g''_2/(2\pi) = 8.51$ MHz | $g'''_2/(2\pi) = 60.2$ MHz |

The operational fidelity is evaluated by

$$\mathcal{F} = \sqrt{\langle\psi_{id}|\rho|\psi_{id}\rangle}, \quad (17)$$

where $|\psi_{id}\rangle$ is the ideal state in Eq. (13), while $\rho$ is the real density operator achieved by numerically solving the master equation (16).

Consider the parameters listed in Table I. The coupling constant $g_2$ is calculated according to Eq. (10) and for $k = 5$ (the optimal value obtained by a numerical test). The $|g'\rangle \leftrightarrow |g\rangle$ and $|g\rangle \leftrightarrow |e\rangle$ transitions are much weaker because of the barrier between the two potential wells. By a proper design of the flux ququart,[47,48] one can have $\phi_{fg'} \sim \phi_{fg} \sim \phi_{fe} \sim \sqrt{2}\phi_{eg'} \sim 10\phi_{eg} \sim 50\phi_{gg'}$, where $\phi_{ij}$ is the dipole coupling matrix element between the two levels $|i\rangle$ and $|j\rangle$ with $ij \in \{fg, fe, eg', eg, gg'\}$. Thus, one has $g'_1 \sim g_1$, $g''_1 \sim 0.1g_1$, and $g'''_1 \sim g_1/\sqrt{2}$ and $g'_2 \sim g_2$, $g''_2 \sim 0.1g_2$, and $g'''_2 \sim g_2/\sqrt{2}$. The maximal coupling constant is $g_{max} = 2\pi \times 95.7$ MHz (Table I), readily available since a coupling constant $\sim 2\pi \times 636$ MHz was reported for a flux device coupled to a microwave cavity.[49]

Other parameters used in the numerical simulations are (i) $\gamma^{-1}_{gg'} = 80$ $\mu$s, $\gamma^{-1}_{eg} = 40$ $\mu$s; (ii) $\gamma^{-1}_{fg'} = \gamma^{-1}_{fg} = \gamma^{-1}_{fe} = \gamma^{-1}_{eg'}/2 = 10$ $\mu$s; (iii) $\gamma^{-1}_{f,\varphi} = \gamma^{-1}_{e,\varphi} = \gamma^{-1}_{g,\varphi} = 5$ $\mu$s; (iv) $\kappa_1 = \kappa_2 = \kappa$; and (v) $\alpha = 0.5$. Here, $\gamma^{-1}_{gg'}$ and $\gamma^{-1}_{eg}$ are larger than $\gamma^{-1}_{eg'}, \gamma^{-1}_{fe}, \gamma^{-1}_{fg}$, and $\gamma^{-1}_{fg'}$ because of $\phi_{eg}, \phi_{gg'} \ll \phi_{fg'}, \phi_{fg}, \phi_{fe}, \phi_{eg'}$ due to the barrier between the two potential wells. The reason for choosing $\gamma^{-1}_{fg'} = \gamma^{-1}_{fg} = \gamma^{-1}_{fe} = \gamma^{-1}_{eg'}/2$ is because of $\phi_{fg'} \sim \phi_{fg} \sim \phi_{fe} \sim \sqrt{2}\phi_{eg'}$. The dephasing times for the three excited levels $|g\rangle, |e\rangle$, and $|f\rangle$ are not associated with the dipole matrix elements and are on the same order. Thus, we choose $\gamma^{-1}_{f,\varphi} = \gamma^{-1}_{e,\varphi} = \gamma^{-1}_{g,\varphi}$. Note that the above choice for the decoherence times is reasonable because an energy relaxation time $T_1 = 60 - 90$ $\mu$s and a dephasing time $T_2 = 40 - 85$ $\mu$s have been experimentally demonstrated for a superconducting flux device.[36,37,50]

Figure 4 gives the fidelity vs $\kappa^{-1}$ for $g_{12} = 0$, $0.01g_{max}$, and $0.1g_{max}$. Figure 4 shows that the fidelity is greater than 98.1% for $\kappa^{-1} \geq 10$ $\mu$s and $g_{12} = 0.1g_{max}$. The setting $g_{12} \leq 0.1g_{max}$ is achievable in experiments by a prior design of the sample with appropriate capacitances $C_1$ and $C_2$ depicted in Fig. 2.[51]

With the parameters listed in Table I, a simple calculation gives $\chi \sim 2\pi \times 1.19$ MHz, resulting in the operational time $t = \pi/\chi \sim 0.42$ $\mu$s. For the cavity frequencies in Table I and $\kappa^{-1} = 10$ $\mu$s, the quality factors of the two cavities are $Q_1 \sim 1.15 \times 10^6$ and $Q_2 \sim 7.03 \times 10^5$, which are available because a 1D microwave cavity with a high quality factor $Q \gtrsim 2.7 \times 10^6$ was reported in experiments.[52,53] The above analysis demonstrates that high-fidelity generation of the cat–cat–spin hybrid GHZ state is feasible with current circuit QED experiments.

As a final note, we investigate the average gate fidelity when each photonic qubit is encoded via the cat states, and each qubit is initially in an arbitrary quantum state. Our numerical simulation demonstrates that the hybrid CCZ gate can be implemented with a high fidelity. For a detailed discussion, see the supplementary material.

See the supplementary material for general effective Hamiltonian, derivation of the Hamiltonian (4), derivation of the Hamiltonian (5), and the average gate fidelity.

This work was partly supported by the National Natural Science Foundation of China (NSFC) (Nos. 11074062, 11374083, 11774076, and U21A20436), the Jiangxi Natural Science Foundation (No. 20192ACBL20051), the Key-Area Research and Development Program of GuangDong Province (No. 2018B030326001), the Jiangsu Funding Program for Excellent Postdoctoral Talent, and the Innovation Program for Quantum Science and Technology (No. 2021ZD0301705).

## AUTHOR DECLARATIONS
### Conflict of Interest
The authors have no conflicts to disclose.

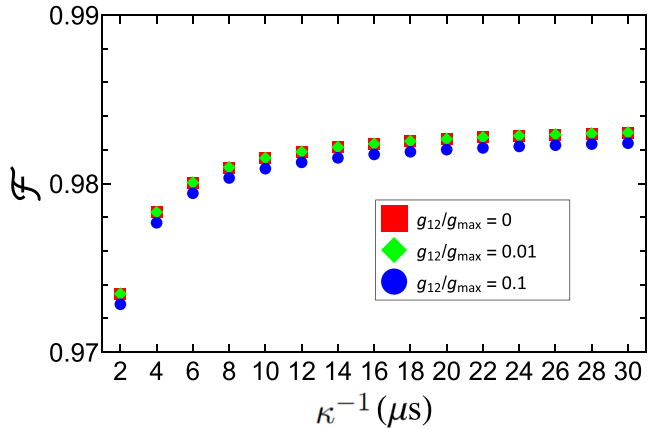

**FIG. 4.** Fidelity vs $\kappa^{-1}$ for $g_{12}/g_{max} = 0$, 0.01, and 0.1. Here, $\kappa$ is the cavity decay rate, and $g_{12}$ is the inter-cavity crosstalk strength.

## Author Contributions

Qi-Ping Su and Yu Zhang contributed equally to this work.

**Qiping Su:** Software (lead); Writing – review & editing (equal). **Liang Bin:** Formal analysis (equal); Writing – review & editing (equal). **Yu Zhang:** Software (supporting); Writing – review & editing (equal). **Chui-Ping Yang:** Conceptualization (lead); Supervision (lead); Writing – original draft (lead).

## DATA AVAILABILITY

The data that support the findings of this study are available from the corresponding author upon reasonable request.